\documentclass[12pt,preprint]{aastex}

\shorttitle{Heating and Dynamics of Two Flare Loop Systems Observed by AIA and EIS}

\shortauthors{Li, Qiu, \& Ding}

\begin{document}

\title{Heating and Dynamics of Two Flare Loop Systems \\Observed by AIA and EIS}

\author{Y. Li$^{1,2,3}$, J. Qiu$^{2}$, M. D. Ding$^{1,3}$}
\affil{$^1$School of Astronomy and Space Science, Nanjing University, Nanjing 210093, China}
\affil{$^2$Department of Physics, Montana State University, Bozeman, MT 59717, USA}
\affil{$^3$Key Laboratory for Modern Astronomy and Astrophysics (Nanjing University), Ministry of Education, Nanjing 210093, China}
\email{yingli@nju.edu.cn}

\begin{abstract}

We investigate heating and evolution of flare loops in a C4.7 two-ribbon flare on 2011 February 13. From {\it SDO}/AIA imaging observations, we can identify two sets of loops. {\it Hinode}/EIS spectroscopic observations reveal blueshifts at the feet of both sets of loops. The evolution and dynamics of the two sets are quite different. The first set of loops exhibits blueshifts for about 25 minutes followed by redshifts, while the second set shows stronger blueshifts, which are maintained for about one hour. The UV 1600 observation by AIA also shows that the feet of the second set of loops brighten twice. These suggest that continuous heating may be present in the second set of loops. We use spatially resolved UV light curves to infer heating rates in the few tens of individual loops comprising the two loop systems. With these heating rates, we then compute plasma evolution in these loops with the ``enthalpy-based thermal evolution of loops'' (EBTEL) model. The results show that, for the first set of loops, the synthetic EUV light curves from the model compare favorably with the observed light curves in six AIA channels and eight EIS spectral lines, and the computed mean enthalpy flow velocities also agree with the Doppler shift measurements by EIS. For the second set of loops modeled with twice-heating, there are some discrepancies between modeled and observed EUV light curves in low-temperature bands, and the model does not fully produce the prolonged blueshift signatures as observed. We discuss possible causes for the discrepancies.
\end{abstract}

\keywords{Sun: corona -- Sun: flares -- Sun: UV radiation}

\section{Introduction}

Solar flares are energetic events in the solar atmosphere. Energy release in flares leads to plasma heating, particle acceleration, and mass motions. During a solar flare, the plasma is heated and fills up the coronal loops. The loops then evolve responding to this energy input and radiate emissions that can be observed. The format of this released energy used to heat flares is still an issue of debate: do non-thermal particles carry a significant amount of energy and deposit most of it in the lower atmosphere and in turn heat the corona? Or does heating take place primarily in the corona and thermal conduction transfers heat flux into the lower atmosphere? And in this process, how much energy is used to heat flare plasmas? These are important but unresolved problems. 

Many coronal heating models have been developed to study how released energy would heat flare plasmas resulting in observed radiation signatures. Modelers usually use an empirical heating function to describe the flare heating. The heating function quantitatively depicts when and where a flare loop is heated, as well as for how long and by how much the loop heated. It is a function of time and space. In time, the heating rate is usually assumed as a single Gaussian profile during the impulsive phase in numerical simulations (e.g., \citealt{fish90, raft09}). It has become recognized that heating may occur continuously throughout the life of a flare, starting from the pre-flare phase, and extending into the decay phase (e.g., \citealt{pall75,born85,phil05}). \cite{batt09} found that the coronal temperature stays constant or even increases during the pre-flare phase. They claimed that continuous heating in the corona is necessary to sustain the observed temperature. In addition, \cite{carg83} proposed that the post-flare loop could be heated by magnetohydrodynamic shocks for a long time. \cite{jian06} reported that continuous energy input is necessary to balance the rapid radiative cooling for a few flares. Continuous heating is also supported by spectroscopic velocity measurements. \cite{czay99} found that the outer edge of the flare ribbon shows blueshifts  in the decay phase, which indicate chromospheric evaporation responding to energy input. These observations provide evidence of continuous heating throughout the lifetime of the flare. In space, the heating function has a distribution along the flare loop. Furthermore, high-resolution imaging observations of the solar corona in the past two decades have led to important progress in our knowledge of flare plasmas: the flare plasmas are confined in numerous loops, whose basic scale is most likely below 1\arcsec~\citep{broo12,pete13,schm13}. These loops are usually formed and heated at different times and evolve independently of one another (\citealt{hori98,asch05,mulu11}). In these different loops the heating rates are most likely different. The concept of heating multiple loops has been applied in recent numerical studies \citep[e.g.][]{warr06,hock12}. These models assume successive heating in newly formed loops to reproduce the observed coronal emission in the decay phase of the flare.

There are several ways to derive the empirical heating function of flares. In some previous studies, a heating rate linearly scaled, by a free parameter, to the observed hard X-ray emission was used to model flare evolution \citep[e.g.][]{fish90}. In other studies that model multiple loop heating, the heating rates in these loops were prescribed by a number of free parameters that are adjusted to match model-produced coronal emission with observations \citep[e.g.][]{hock12}. These models are therefore constrained by comparing the model output---the computed coronal loop emission, with observations. Recently, \cite{long10} computed a heating rate using a reconnection model with measured reconnection flux in flares, and predicted the coronal radiation to compare with observations. \cite{qiuj12} presented a new method to derive the flare heating rates using spatially resolved ultraviolet (UV) light curves, and then compute plasma evolution in the heated loops to compare with observations. This recent method, compared with earlier ones, constrains the heating rates from both the input and output.

In this paper, we study heating of two sets of flare loops (or two loop systems) identified from imaging observations. Each loop system is considered to be composed of a bundle of thin loops with their size given by the instrument pixel scale. These loops connect opposite magnetic polarities. We adopt the method of \cite{qiuj12}, with the heating rate observationally inferred from spatially resolved UV light curves at the footpoints of flare loops that are heated independently. Since the transition region responds rapidly to energy deposition and generates enhanced UV emission with a timescale much shorter than the observational cadence \citep{canf87,fish85}, the observed impulsive rise of the UV light curve indicates the starting time and duration of heating in each loop, and also roughly indicates the magnitude of heating---namely, the brighter pixels are assumed to be heated with a greater heating rate. With this method, the heating rates in numerous flare loops can be mostly inferred from UV observations and used to compute plasma evolution in these loops. In this study, there is only one free parameter to scale the peak heating rate to the observed peak UV emission, which is further determined by comparing model outputs with coronal observations \citep{qiuj12,ying12,liuw13}. Therefore, parameters of the heating rates (time, duration, and relative magnitude) for individual loops are constrained from observations, and this data-driven flare heating function is temporally and spatially resolved.   

We use these heating rates to compute plasma evolution in flare loops using the zero-dimensional (0D) model called ``enthalpy-based thermal evolution of loops" (EBTEL; \citealt{klim08, carg12a}) in this study. The EBTEL model computes evolution of the average temperature and density in a coronal loop given an impulsive energy input, which has been validated by advanced one-dimensional (1D) hydrodynamic simulations. Although the model works under very simplified assumptions, its high efficiency and very few free parameters allow us to calculate mean plasma properties for a large number of flare loops. As these loops are heated at different times and evolve independently, at any given time, these different loops are in different evolution stages with different mean temperatures and densities. Therefore, this method naturally produces the differential emission measure as the temperature distribution of the plasma contained in the large number of loops, and allows us to calculate the total radiation in different wavelengths to compare with observations. There is another reason to use this model: at present we do not clearly know how the heating rate distributes along the flare loop from observations. Therefore, the 0D EBTEL model may be a sensible choice to compute plasma evolution to the first order. Furthermore, in this study, the number of available observables is much greater than the number of free parameters used in the model, hence comparison of the model with observed flare radiation and dynamic properties will provide a test of the model assumptions.

Through the model, the synthetic extreme-ultraviolet (EUV) flux (or intensity) and mean enthalpy flow velocity can be computed and then compared with observations. The presented observations here are high-resolution images from the Atmospheric Imaging Assembly (AIA; \citealt{leme12}) on board the {\it Solar Dynamics Observatory} ({\it SDO}) and spectra from the EUV Imaging Spectrometer (EIS; \citealt{culh07}) on board {\it Hinode}. Due to its observing mode, it takes some luck for EIS to capture a flare. Here, fortunately, EIS observes a flaring region with a good temporal resolution. It provides the information of flare radiation and dynamics as well as their evolution. EIS observations of this flare reveal some new results. AIA high tempo-spatial images provide more complete information on flare evolution and fine structure, which help constrain the model from both the input (heating function) and output (coronal radiation). Such observational analysis combining with modeling provides better diagnostics of flare heating and dynamics. 

\section{Observations}
\label{observation}

\begin{figure*}
\begin{center}
\includegraphics[width=13cm]{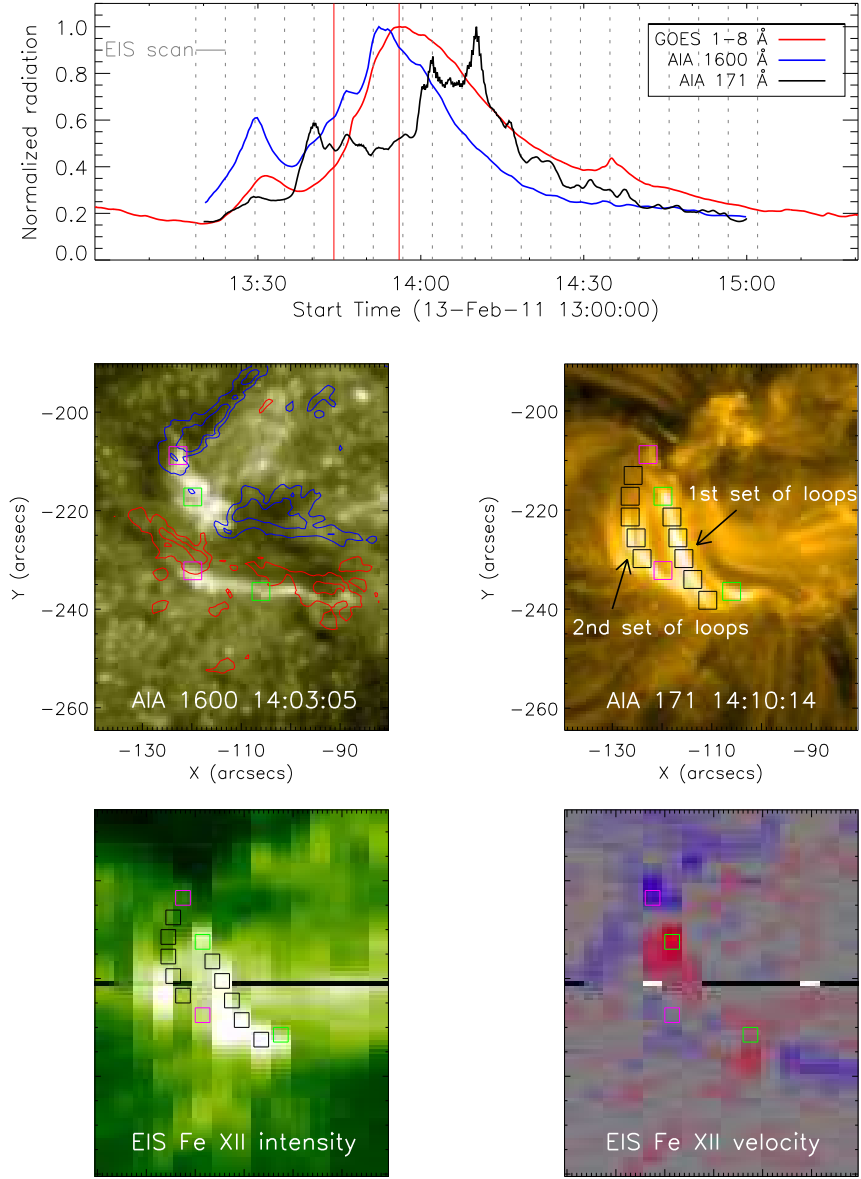}
\caption{Observations of the C4.7 two-ribbon flare that occurred on 2011 February 13. The top panel plots the normalized light curves in different wavelengths. The two vertical red lines represent the times of the flare onset and {\it GOES} soft X-ray peak. The vertical dashed lines mark the 19 EIS scans covering the flare time. The middle panel shows the images in AIA 1600 \AA~and 171 \AA~bands. The red and blue contours show the positive and negative magnetic polarities respectively. The two pairs of color boxes (green and pink) denote two sets of conjugate footpoint patches, and the series of black boxes are used to track the loop segments, with each box 4\arcsec~by 4\arcsec. The bottom panels plot the intensity and Doppler velocity maps in the EIS Fe {\sc xii} 192.39 \AA~line at $\sim$14:10 UT. All the four images have the same field-of-view.}
\label{fig_obs}
\end{center}
\end{figure*}

The flare analyzed here is a C4.7 two-ribbon flare that occurred on 2011 February 13. Figure \ref{fig_obs} presents the observations. The top panel plots the normalized light curves in {\it GOES} soft X-ray 1--8 \AA, AIA UV 1600 \AA~and EUV 171 \AA~wavelengths. All three light curves exhibit two major peaks. In time sequence, the UV emission rises first, followed by the soft X-ray and then EUV emissions. This indicates heating followed by cooling of coronal plasmas. The middle panel shows the images in the UV and EUV bands from AIA. We can see the two ribbons and post-flare loops clearly. The bottom panel gives the intensity and Doppler velocity maps in the EIS Fe {\sc xii} 192.39 \AA~line. The velocity map reveals some upflows and downflows in the flare loops. For this event, although {\it RHESSI} and {\it Fermi} both show enhanced emission of up to 25 keV in the first few minutes as well as in the decay phase, there is no hard X-ray observation during most of the flare period because of {\it RHESSI} night. Therefore, there is no strong evidence that thick-target non-thermal emission be significant during the flare.

AIA observes the UV and EUV images with the pixel scale of $0.\!\!^{\prime\prime}6$ by $0.\!\!^{\prime\prime}6$ and time cadences of 24 s and 12 s respectively. Such high tempo-spatial resolution allows us to identify the flare loops and their footpoints. EIS scanned over the flaring region using the 2\arcsec~slit with a 4\arcsec~step mode. The spatial resolution is $4\arcsec$ by 1$\arcsec$ in the directions across the slit and along the slit respectively. It took EIS 5 min and 27 s to scan over the region once, and the 19 scans covering the flare period are marked by the vertical dashed lines in Figure \ref{fig_obs}. The EIS data are corrected for dark current, hot pixels, and cosmic ray hits by using the standard reduction package. The tilt of the EIS slit is also corrected. In addition, we correct the variations of spectral line positions throughout the orbit by averaging the line centers over a relatively quiet region (the bottom 50 rows in EIS raster). In this study, we select eight EIS spectral lines, listed in Table \ref{tab_line}, to study the flare emission and dynamics. All the line profiles, which exhibit a good symmetry, are fitted by a single Gaussian function with constant background. When measuring the Doppler velocity, we average the line center over the quiet region as the reference wavelength. Here, only the unblended lines are chosen to determine the Doppler velocity. The co-alignment between AIA and EIS is achieved by comparing the 193 \AA~image and Fe {\sc xii} 195.12 \AA~intensity map.

\begin{table}
\begin{center}
\small
\caption{Spectral Lines Used in This Study}
\label{tab_line}
\begin{tabular}{lcc}
\tableline
\tableline
\multicolumn{1}{l}{Ion}	&$\lambda$(\AA) &$T_{max}$(MK) \\
\tableline
  Fe {\sc x}	&184.54	&1.0\\
  Fe {\sc xii}	&192.39	&1.25\\
  Fe {\sc xiii}	&202.04   &1.5\\
  Fe {\sc xv}	&284.16	&2.0\\
  Fe {\sc xvi}     &262.98   &2.5\\
  Ca {\sc xiv}    &193.87	&3.2	\\
  Ca {\sc xvii}   &192.82	&5.0	\\
  Fe {\sc xxiii}	&263.76	&12.5\\
\tableline
\normalsize
\end{tabular}
\end{center}
\end{table}

\subsection{Velocities and Light Curves at Footpoints of the Two Flare Loop Systems}

\begin{figure*}
\begin{center}
\includegraphics[width=16cm]{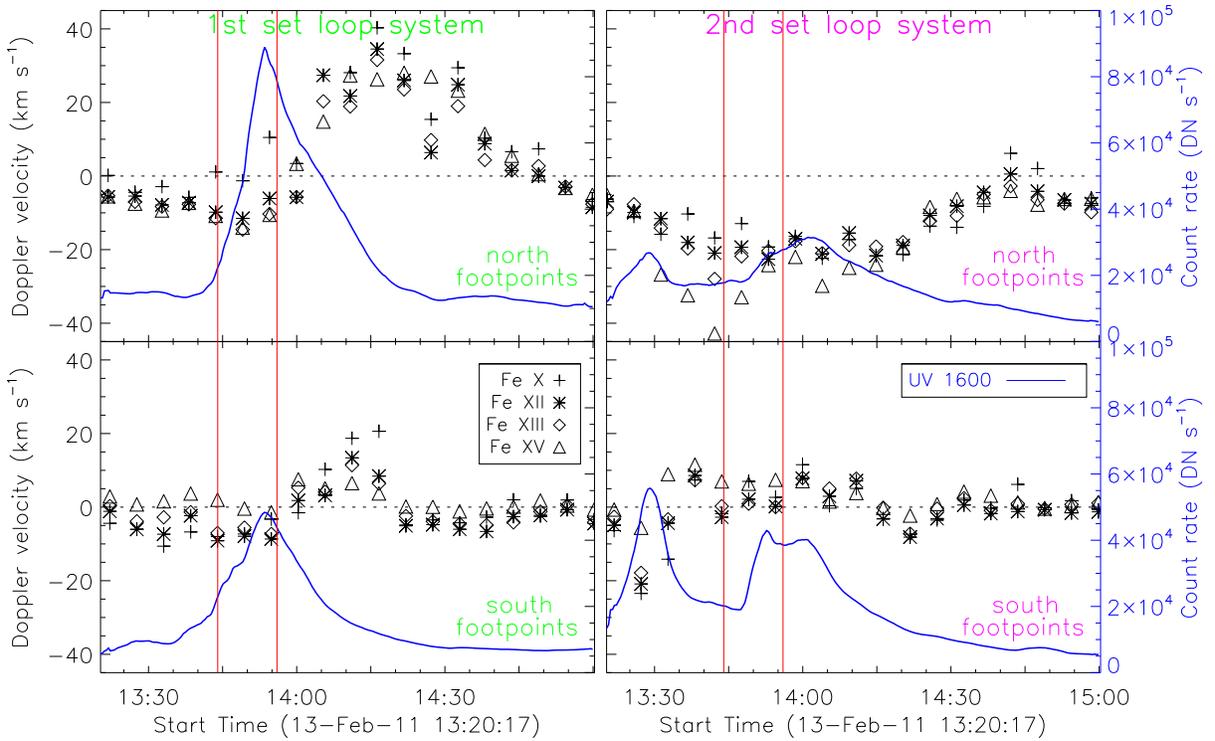}
\caption{Averaged Doppler velocities (black symbols; corresponding to the left coordinate) in four EUV lines and integrated light curves (blue lines; corresponding to the right coordinate) in UV 1600 \AA~at the four footpoint patches of the two loop systems. The negative velocity represents blueshift. The two vertical red lines in each panel mark the times of the flare onset and {\it GOES} soft X-ray peak, the same as in Figure \ref{fig_obs}.}
\label{fig_fp}
\end{center}
\end{figure*}

Based on the flare morphology in EUV images and the UV light curves at conjugate footpoints, we identify two flare loop systems. Their footpoints are marked by the two pairs of color boxes (green and pink) with a cross-sectional area of 4$\arcsec$ by 4$\arcsec$ (limited by EIS spatial resolution) in Figure \ref{fig_obs}. The loop segments are tracked by a series of black boxes (each also with 4$\arcsec$ by 4$\arcsec$). Hereafter we call the loop system located in the west (with footpoints plotted in green) as the first set of loops, and call the other one in the east (with footpoints plotted in pink) as the second set. These two sets of loops show very different behaviors in EUV Doppler velocity and also UV light curve. Figure \ref{fig_fp} plots the averaged Doppler velocities in multiple EUV spectral lines at the four footpoint patches, together with the integrated light curves in UV 1600 \AA. It is seen that the north footpoints of the first set of loops exhibit blueshifts (negative velocities) of $\sim$10 km s$^{-1}$ for about 25 minutes, followed by redshifts (positive velocities) of greater than 20 km s$^{-1}$. For the second set of loops, the north footpoints show stronger blueshifts of $\sim$20 km s$^{-1}$, which are maintained for about 1 hour. Note that, for both sets of loops, the intensity light curves are very similar at the northern and southern foot patches; however, the measured Doppler velocities at the south footpoints are significantly smaller than at the north footpoints, particularly in the second loop system. This might be mainly due to the projection effect: first, different viewing angles of conjugate footpoints lead to different line-of-sight velocity components; second, there might be ambiguity in the velocities measured at the south footpoints, since these measurements may include flows at the footpoints as well as part of the loops that appear to overlap the footpoints. Therefore, we only discuss the Doppler velocities at the north footpoints of the two loop systems. According to the chromospheric evaporation theory \citep{hira74,acto82,anto84,czay99}, the observed long-lasting blueshifts in the second set of loops suggest that continuous heating may be present for a long time. This is also evident in the UV 1600 \AA~light curves at the footpoints of the loops. The light curves of the second set of loops show two peaks before and after the flare onset; while the light curves of the first set just brighten once during the rise of the soft X-ray flux. The prolonged blueshifts observed at the footpoints of the second set of loops are coincident with the UV brightening.

The AIA UV 1600 \AA~channel contains continuum and the C {\sc iv} line. During a flare, the emission in this channel is significantly enhanced, with the excess emission originating mainly from the upper chromosphere or transition region. It is the most direct evidence of energy release. The rapid rise of the UV emission reflects the energy deposition into the lower atmosphere. In this flare, the different evolutions of UV light curves of the two loop systems indicate different cases of energy release (or heating): the energy release in the first set of loops occurs during the impulsive phase, consistent with the traditional point of view; while in the second set of loops, the energy is continuously released during the flare, starting from the pre-flare phase, and extending to the decay phase. This continuous energy release is also reflected by the Doppler velocity measurements at the footpoints of the second set of loops, which show long-lasting blueshifts. These observations suggest that the time history of energy release is different in these two sets of loops adjacent to each other.

\section{Modeling of the Two Sets of Flare Loops}
\label{modeling}

The observations show that the heating process of this flare is very complicated, varying with different loops and spanning different phases of the flare. Measurements of the Doppler shifts at the footpoints are averaged in the 4\arcsec~by 4\arcsec~area (four EIS pixels) due to the coarse spatial resolution of EIS; however, these footpoint patches can be better resolved in AIA images with a pixel scale of $0.\!\!^{\prime\prime}6$, which means that there are 49 AIA pixels in a footpoint patch (a green or pink box in Figure \ref{fig_obs}). We consider that anchored to each AIA pixel is a flare loop with the cross-sectional area of $0.\!\!^{\prime\prime}6$ by $0.\!\!^{\prime\prime}6$; therefore each loop system is composed of such 49 individual loops. The spatially resolved UV light curves observed by AIA can provide information of the heating history at the footpoints of these 49 loops. We adopt the method of \cite{qiuj12} to observationally infer heating rates from UV light curves at the footpoints of these loops, and model plasma evolution with a 0D loop heating model, the EBTEL model \citep{carg12a}. We then synthesize coronal radiation and enthalpy flow velocity to compare with EUV observations. 

The EBTEL model describes the response of plasmas in the coronal loop to a given impulsive heating function. The model solves an energy equation governed by the input heating rate and total loss rate. The net loss includes the optically-thin radiative loss in the corona and an amount of loss through the transition region, part of which will be radiated by the chromosphere. The loss through the transition region is either directly scaled to the coronal radiative loss with a free scaling parameter \citep{klim08}, or estimated by taking into account the atmosphere stratification \citep{carg12a}. In this study, we have computed loop evolution with both versions of the code, and found that the difference is only about 10\% in both the temperature and density values. In this paper, only the results computed using the second version of the model \citep{carg12a} are presented. The EBTEL model also solves a mass equation by considering the flow across the boundary between the corona and its base, or the transition region; this flow is computed in the form of the enthalpy flow as the difference between the conductive heat flux into the transition region and the loss from this region, when non-thermal particle precipitation is insignificant. In this study, lacking hard X-ray observations to identify the presence of non-thermal particles, we use the EBTEL code to model flare loops primarily heated in the corona, and energy flux is transferred to the transition region primarily by conduction. The model uses very few parameters; it is highly efficient and particularly suitable for computing mean plasma properties in a large number of loops and deriving the sum radiation of these loops. Note that EBTEL models a half loop assuming symmetric heating. Seen from Figure \ref{fig_fp}, conjugate footpoints of each set of flare loops exhibit rather similar UV light curves, hence we may justify the symmetry assumption, and model 98 half loops anchored at 98 AIA pixels for each of the two loop systems.

\begin{figure*}
\begin{center}
\includegraphics[width=15cm]{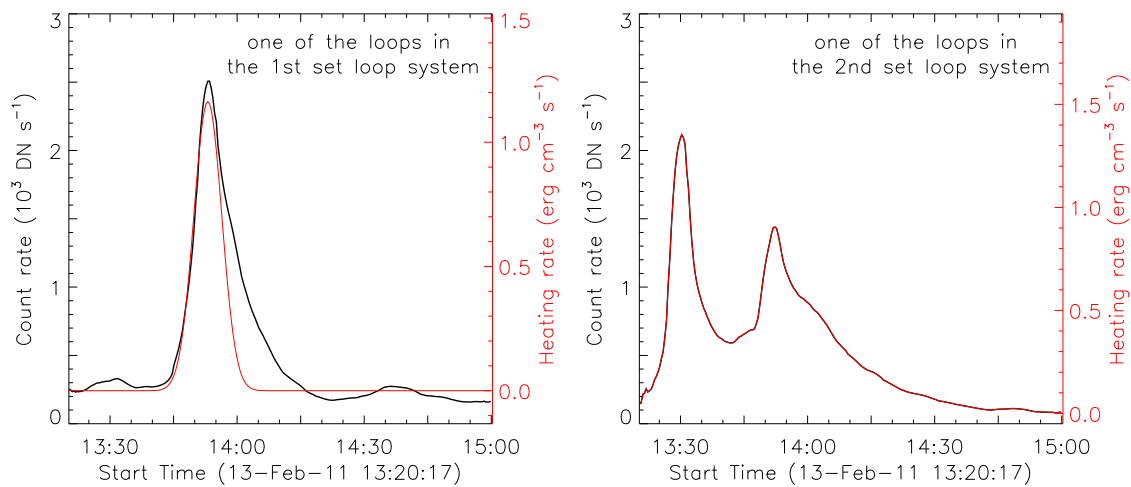}
\caption{Left panel: heating profile (red line) of one loop in the first loop system. It is derived from Gaussian fitting to the observed UV light curve (black line). Right panel: heating profile of one loop in the second loop system, which is entirely proportional to the UV light curve.}
\label{fig_hp}
\end{center}
\end{figure*}

The inputs of the EBTEL model are half-loop length $L$ and volumetric heating rate $Q(t)$. They are derived from observations in our study. For this flare, the half lengths of these two sets of loops are both about 13 Mm estimated from AIA images with a semi-circle assumption. The heating rate of a half loop is inferred from the UV light curve at the pixel where the loop is anchored. It is expressed as $Q(t) = \lambda\,H(t)/L$, where $H(t)$ is a shape function determined from the observed UV count rate at this pixel, $\lambda$ is a scaling constant, which is the same for all 98 half loops in each set, that can be determined by best-matching modeled and observed total coronal radiation from the set of loops. Figure \ref{fig_hp} shows the UV light curves from two of the half loops, each from a loop system. All 98 loops in the first set exhibit a single peak at slightly different times during the impulsive phase, whereas about one half of the second set of loops exhibit two peaks as shown in the right panel of the figure, and the other half show only one peak in the UV light curve. It suggests that the first set of loops are heated once, but twice-heating is present in one half of the second set of loops. The $H(t)$ function used here is different in these two sets of loops. In the first set, it is assumed to be a full Gaussian derived by fitting the rise phase of the footpoint UV 1600 \AA~light curve to a half Gaussian. Note that, we do not use the decay phase to infer the heating rate for this set, since the decay phase is not a direct reflection of energy release but mostly a manifestation of plasma cooling. While for the second set of loops, due to the complexity in the heating history, and more importantly, because we wish to examine whether significant continuous heating is present in the decay phase so as to produce long-lasting blueshifts, $H(t)$ is assumed to be the observed UV light curve itself (its effects on radiation will be discussed below). The finally derived heating rates (red lines in Figure \ref{fig_hp}) in these loops typically range from 0.3 to 1.3 ${\rm erg\ cm}^{-3}~{\rm s}^{-1}$. 

\begin{figure*}
\begin{center}
\includegraphics[width=15cm]{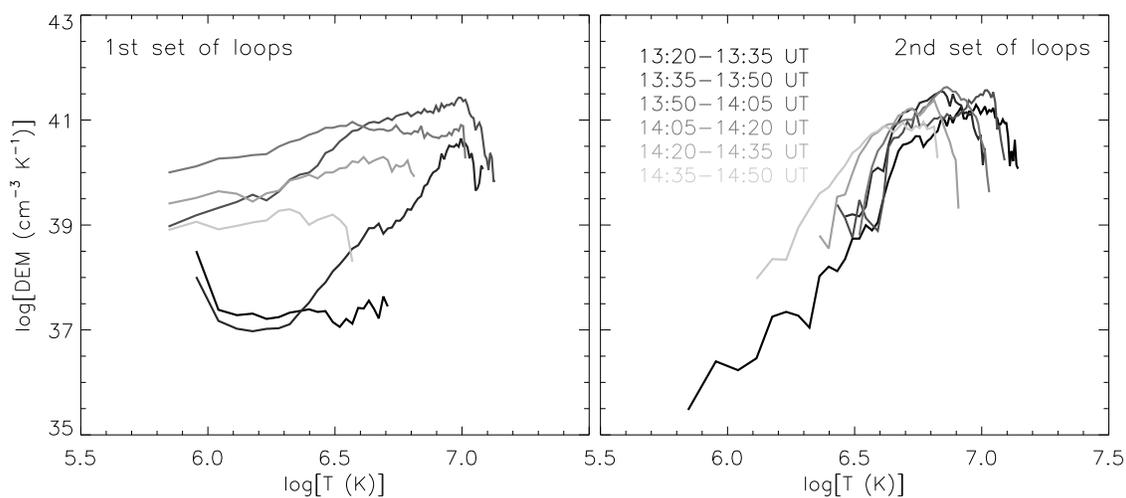}
\caption{Coronal DEM of the two sets of loops averaged every 15 minutes from 13:20 UT (dark to grey).}
\label{fig_ev}
\end{center}
\end{figure*}

Using the inputs described above in the EBTEL model, we compute plasma evolution for each half loop. As these loops are heated at different times and presumably by different amounts of energy, they evolve independently of one another, and exhibit different mean temperatures and densities at any given time. The temperatures and densities of these 98 half loops are used to compute the differential emission measure (DEM) of the coronal plasma for each set of loops with this formula: $DEM=n^2\ dV/dT$, where $n$ is the electron density, $V$ is the total volume of the plasma, and $T$ is the temperature; in practical calculations, $DEM(i) = (A \sum_{j = 1}^{N_i} n_j^2 L_j)/\Delta T(i)$, where $\Delta T (i) = T(i+1) - T(i)$, $A$ is the loop cross-sectional area, $j$ indicates a loop that has a temperature falling in the range between $T(i)$ and $T(i+1)$, and $N_i$ is the number of this subset of loops. Figure \ref{fig_ev} shows the coronal DEM for the two sets of loops averaged every 15 minutes during different evolution stages of the flare. For the first set of loops, it is seen that the DEM at high temperatures is not significant in the early stage (13:20--13:35 UT). However, it becomes significant with the heating added (13:35--14:05 UT). In the late evolution stage (after 14:05 UT), the DEM decreases towards the low temperatures, which reflects cooling of plasmas. For the second set of loops, the DEM distributes in a wide temperature range in the early stage, and changes slowly at high temperatures during the subsequent evolution stages. That is due to continuous heating in this set of loops.

\section{Comparison between Modeling and Observations}
\label{comparison}

From the computed temperatures and densities, we can derive synthetic coronal emission fluxes of the loops. For each loop system, the fluxes from the 98 half loops are added up, and then compared with the EUV emission observed by AIA and EIS. The observed emission is measured from a series of 4\arcsec~by 4\arcsec~boxes along the loop system (see Figure \ref{fig_obs}). The model also computes the enthalpy flow velocity at the coronal base, which then can be compared with the Doppler velocity measured by EIS at the footpoints of the loop system. 

\subsection{Comparison of EUV Fluxes}

\begin{figure*}
\begin{center}
\includegraphics[width=15cm]{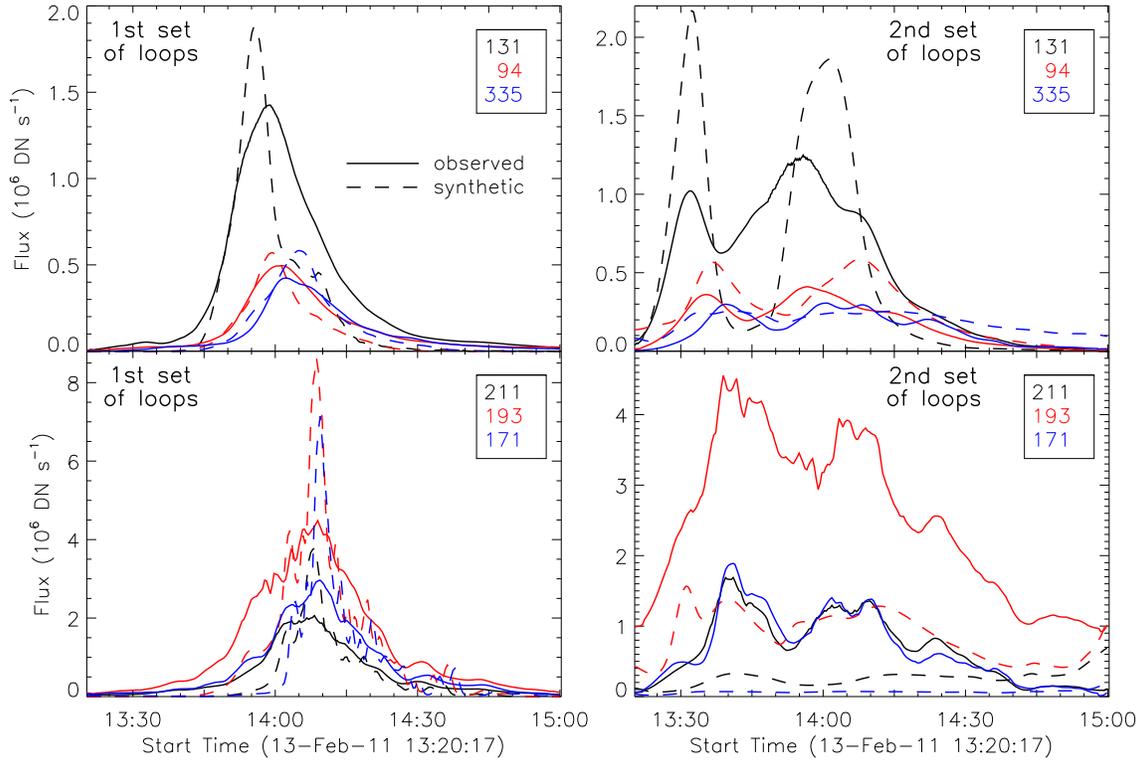}
\caption{Comparison between synthetic (dashed lines) and observed (solid lines) EUV fluxes of the two sets of loops in six AIA channels. The numerous fluctuations in the observed 171 \AA, 193 \AA, and 211 \AA~curves are caused by saturation in these channels.}
\label{fig_comAIA}
\end{center}
\end{figure*}

Firstly we compare the synthetic EUV fluxes with AIA observations. Figure \ref{fig_comAIA} shows the comparison in six EUV channels. Note that there is saturation in some channels, e.g., 211 \AA, 193 \AA, and 171 \AA. So we adopt the fluxes at short exposure times for these channels with smaller saturation effect. From the figure it is seen that,  for the first set of loops, the computed EUV fluxes generally agree with the observed fluxes in the six AIA channels. They all exhibit similar evolution trends, as well as rise and reach peak at nearly the same time; their magnitudes are also comparable. There are discrepancies in a few places. In the 131 band which observes plasmas at high temperature ($\sim$10 MK), the modeled flux rises and decays somewhat faster than the observed. And in the low-temperature bands (211, 193, and 171), the model-computed flux rises a bit later than the observed, and exhibits a surge at around 14:10 UT, when the observed flux also peaks but with a smaller magnitude. For the second set, it is shown that continuous heating in the model can produce double EUV emission peaks as observed in all AIA bands. However, compared with the first set of loops, the model generates more flux in high temperature bands (131 and 94), which are sensitive to plasmas at temperatures of 6--10 MK, and insufficient flux in the low temperature bands (211, 193, and 171) sensitive to temperatures of 1--2 MK. In addition, for this set, the model-generated second peak is delayed with respect to the observed peak in a few bands. 

\begin{figure*}
\begin{center}
\includegraphics[width=15cm]{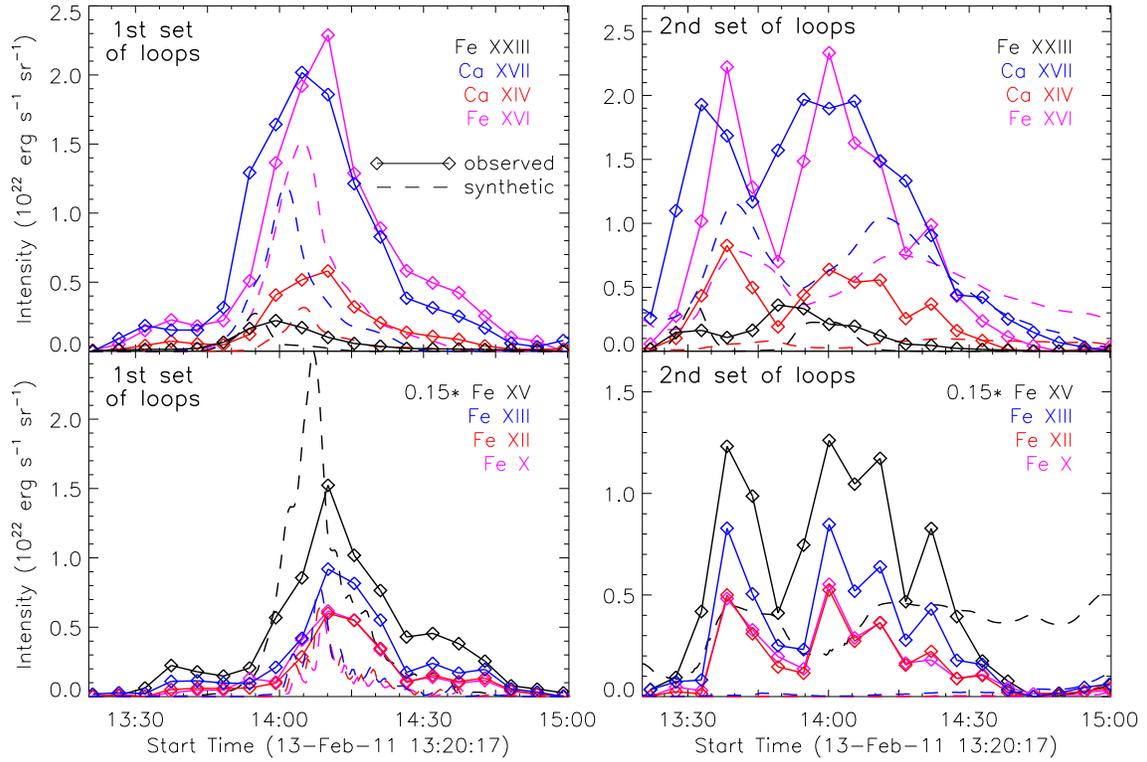}
\caption{Comparison between synthetic (dashed lines) and observed (diamonds with line) intensities of the two loop sets in eight EIS spectral lines. Here, both of the observed and synthetic intensities in the Fe {\sc xv} 284.16 \AA~line are multiplied by 0.15 in oder to display well.}
\label{fig_comEISI}
\end{center}
\end{figure*}

We also compare the model results with EIS observations. Figure \ref{fig_comEISI} shows the observed and computed intensities of the eight EIS lines listed in Table \ref{tab_line} for the two loop systems. For the first set of loops, the computed intensity basically tracks the observed intensity in all eight lines; in a few lines, the model under-estimates the intensity by within a factor of two. For the second set of loops, similar to the AIA comparison, the model results and observations both show double peaks in a few relatively hot lines, though the second peak of the synthetic flux in these lines is somehow delayed by a few minutes compared with that observed. Similar to the AIA comparison, the model generates very little flux in low-temperature Fe {\sc x} 184.54 \AA, Fe {\sc xii} 192.39 \AA, and Fe {\sc xiii} 202.04 \AA~lines. 

In summary, we see that AIA and EIS observations are rather consistent with each other, while there exist some discrepancies between model and observations. The discrepancy may be partly caused by the mean temperature and density approach of the 0D EBTEL model. In reality, there are temperature and density distributions along a single loop during its evolution, which may further broaden the DEM curve and may improve the model-observation comparison. It is also noted that, for the first set of loops, the heating rate is derived as a full Gaussian by fitting only the rise phase of the UV light curve; whereas for the second set of loops, the heating rate is directly scaled to the observed UV light curve from its rise to decay. We have compared the model results for the second set of loops using different kinds of heating rates. Our experiments show that, these two different heating rate profiles have only minor effects on the radiation in high-temperature bands; the heating rate directly scaled to the UV light curve reduces radiation in low-temperature bands, and delays the peak of low-temperature emission by a few minutes. These effects on the low-temperature emission are caused by the extra heating into the decay phase, which leads to slower cooling of plasmas. The computed radiation using this heating rate compares more favorably with the observed EUV light curves than using the heating rate from half-Gaussian fitting, suggesting that continuous heating into the decay phase may not be ignored in the second set of loops. Given the limitation of the model and the fact that only one free parameter, the scaling constant ($\lambda$) for the magnitude of heating rates, is used in modeling all the loops, we regard that there is rather good agreement between model results and observations, particularly for the first set of loops.

\subsection{Comparison of Footpoint Flow Velocities}

\begin{figure*}
\begin{center}
\includegraphics[height=17cm]{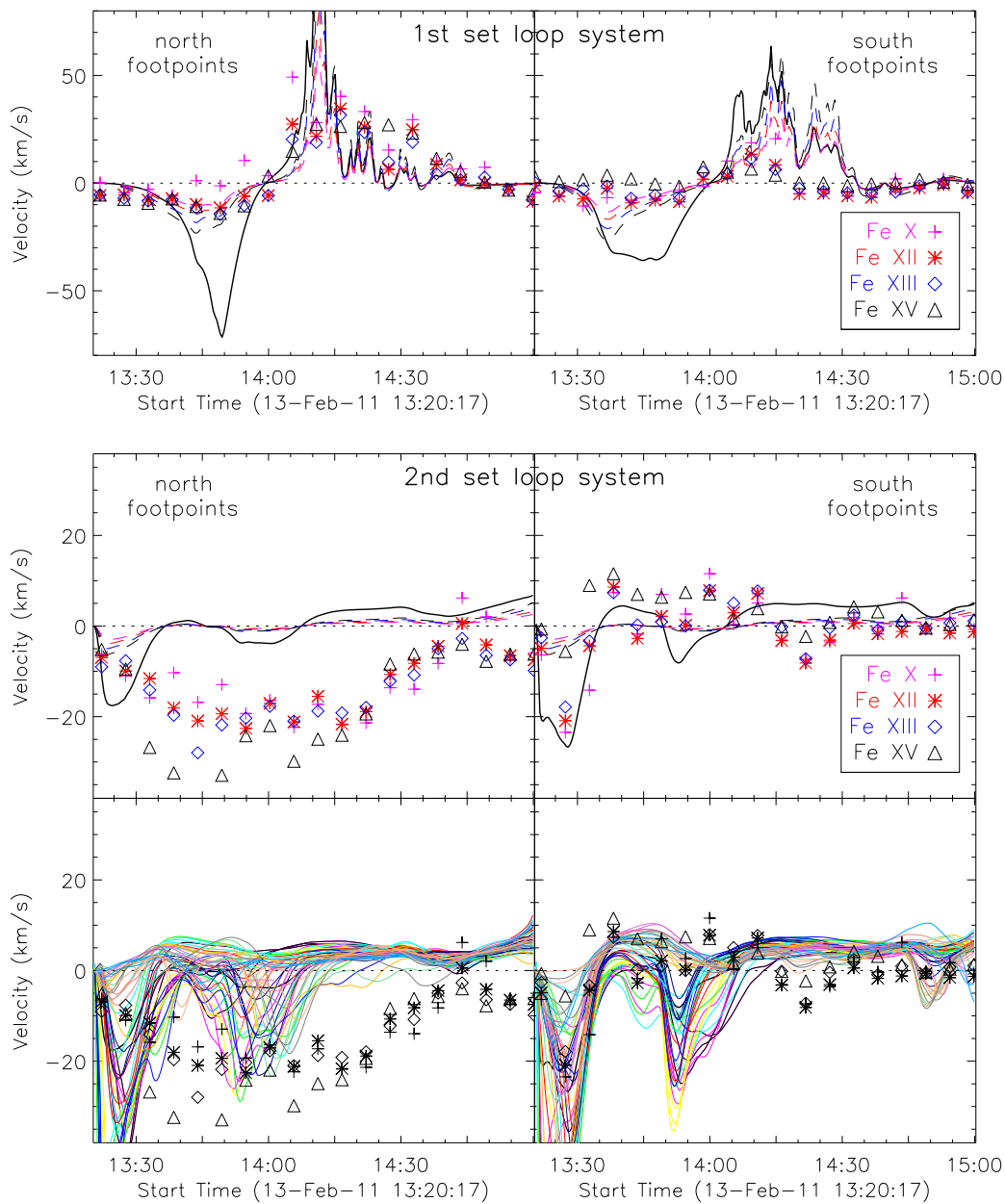}
\caption{Comparison between synthetic velocities (lines) and measured Doppler velocities (symbols) at footpoints of the two sets of loops in multiple EIS lines. The black solid lines in the first four panels indicate the enthalpy flow velocity averaged over the 49 loops, and the dashed lines are the computed equivalent velocities at the formation temperatures of EIS lines. The color lines in the bottom two panels plot the evolution of enthalpy flow velocities for all the 49 loops in the second loop system.}
\label{fig_comEISv}
\end{center}
\end{figure*}

Besides EUV emissions, we also compare model-estimated flow velocity with observations. Note that, the EIS Doppler shift measurements are reliable in the four low-temperature lines, which are unblended and with high signal-to-noise ratios for this event; while there are large uncertainties in the measurements for the other four high-temperature lines. Figure \ref{fig_comEISv} shows the comparison of velocities in the four low-temperature lines. The measured Doppler shifts (in color symbols) are rather homogeneous for all the four lines without apparent temperature dependence. The model computes the flow velocities at the coronal base of 49 half loops for each footpoint patch; the solid dark line in the figure gives the average velocity of these 49 half loops. To estimate the velocity in the EIS lines, we assume uniform pressure and steady-state flow across the coronal base, in which case, the velocity in each line is roughly proportional to the line formation temperature. The colored dashed lines in the figure show the estimated velocities in the four lines at the footpoint patches for the two sets of loops.

For the first set of loops, it appears that the model-estimated velocities compare very well with the EIS measurements, even though the assumption of steady-state equilibrium is rather inaccurate. It is encouraging to note that the model-estimated and observed velocities both switch from upflow to downflow at around the same time. The agreement is most prominent for the north footpoint where the projection effect is less significant. For the second loop system, the synthetic velocities do not match well the measurements. The synthetic flows are weak and short-lived, inconsistent with the observed strong and long-lasting blueshifts. The bottom panel of the figure also shows flow velocities in all the 49 loops. The flow velocities of individual loops are comparable with observations during the times of heating; however, the synthetic flow speed does not match the observed phenomenon. Furthermore, observations show continuous upflows in between the two periods of heating, which cannot be produced by the model.

In summary, the model-observation comparison is different in the two sets of loops. For the first set of loops, given the limitation of the 0D EBTEL model and that only one free parameter is used in modeling nearly one hundred loops, the model results can be considered to be in very good agreement with the observed fluxes and flow signatures measured in multiple EUV lines characteristic of plasma properties at different temperatures during different evolution stages. These results suggest that the 0D multi-loop heating model is able to correctly describe the mean radiation and dynamic properties of this set of loops. We note that EBTEL only solves a gross energy equation without treating the detailed physics during the heating phase, and in this study, we model the loop heating with the assumption that heating primarily takes place in the corona with negligible precipitating non-thermal particles. The agreement between model and observations in this set of loops would largely indicate that the model assumptions, such as heating in the corona and energy transport primarily by conduction, are not far from the real situation, and that in this case, the timing and total amount of heating energy of each of the heating events, which we infer from UV observations, are important for the net radiative output and dynamic evolution of these loops, regardless of the details of the heating phase. For the second set of loops which are considered to be twice-heated, there are large discrepancies between model and observations particularly for the late heating phase and in low-temperature (1-2 MK) lines: the model does not generate enough EUV fluxes in these lines, and also fails to produce the long-lasting strong blueshifts measured in these lines; the model-produced second emission peak is delayed by a few minutes with respect to observations. These may indicate the complexity in plasma evolution with continuous heating not appropriately described by the present method.

\section{Summary and Discussions}
\label{discussion}

We have investigated heating and evolution of two loop systems in a C4.7 two-ribbon flare through AIA imaging and EIS spectroscopic observations and EBTEL modeling of multiple loops. The two sets of loops show quite different behaviors: the first set exhibits 25-minute blueshifts at the footpoints followed by redshifts, while the second set shows blueshifts for about 1 hour indicative of continuous heating. In the mean time, the spatially resolved UV light curves from the feet of the two sets of loops show single and twice brightenings respectively, implying heating with different time scales. With heating functions inferred from the UV light curves at loop footpoints, we model the evolution of flare plasmas in these two sets of loops, compute the coronal radiation and enthalpy flow velocity, and compare them with observations. 

The two loop systems show quite different evolution and dynamics in UV and EUV observations, which demonstrate that there exist different heating processes in these two sets of loops. The first set loops are mainly heated during the impulsive phase of the flare, while the second set loops appear to be heated twice, once during the pre-flare phase, and then again during the decay phase. These are reflected in the footpoint UV emissions, and {\it GOES} soft X-ray light curve also shows a few peaks corresponding to these different episodes of heating in these loops. As a consequence of different heating processes, the two sets of loops exhibit different dynamic behaviors. Spatially resolved observations allow us to further divide these different heating events into heating a few tens of loops, and therefore construct time and space (loop) dependent heating functions for the flare. The evolution and dynamics of the first set of loops are verified by the model that produces emission and flow patterns consistent with observations. The agreement suggests that the EBTEL model, even though with the assumption of uniform heating and only computing mean properties in the loop, captures major features of the loop evolution in this set of loops, which are likely heated primarily in the corona and transfer energy downward by conduction. However, the same modeling cannot satisfactorily reproduce the observed EUV flux and flow signatures in low temperature lines for the second set. These may indicate different heating mechanisms in the two sets of loops that might not be treated by the same method. 

The prominent discrepancies between model results and observations for the second set of loops may be caused by the following reasons. (1) The heating may be largely non-uniform along the loop, likely concentrated on the loop footpoints. This is implied by the limited hard X-ray observations from {\it RHESSI}, which show some bursty emissions up to 25 keV just before and after the flare impulsive phase. Based on this fact, non-thermal electrons may play some role in heating this set of loops. However, such non-uniform heating cannot be modeled by the present 0D EBTEL model. (2) There may exist unresolved fine flaring strands, and the continuous heating takes place in different strands by very different rates as well as at quite different times. The continuous heating in different unresolved strands would produce emissions at high temperatures by plasmas in newly heated strands as well as emissions at low temperatures by earlier heated strands that already cool down to 1-2 MK. Compared to the case of heating continuously occurring in the same resolved loop, this would produce long-lasting outflows as well as enough cool emissions, making the model results more consistent with observations. The fine strands seems to be unresolved by the AIA resolution \citep{broo12}. We need higher resolution observations to infer the heating functions and model the plasma evolution in these fine strands.

Though with simplified assumptions, the 0D EBTEL model uses only one free parameter, and is very fast in computing mean plasma properties. Such high efficiency allows the model to be very useful for a quick-look of loop evolution, for parameter exploration and survey, and particularly for modeling a large number of loops \citep{carg12a,carg12b}. It can constrain the flare heating model, and provide a reference for 1D or multi-dimensional models. Our study shows the advantage of using EBTEL to model a large number of loops with reasonable agreement with observations in many aspects. On the other hand, the presented discrepancies between the 0D model and observations provide insights into how a 1D model may possibly address these discrepancies, improve model-observation comparison, and obtain more accurate diagnostics of flare heating and plasma evolution.

Our spectroscopic analysis provides diagnostics of the flare dynamics. Plasma flow is crucially involved in the process of flare heating and evolution. The velocity measurements, on one hand, can be used to determine free parameters in the model (see \citealt{ying12}); on the other hand, they can help examine the assumptions made in the model, from which we deduce that the enthalpy flow velocity is proportional to temperature along the loop. The linear relation between velocity and temperature along the loop has been tested to be valid in at least some observations (e.g., \citealt{mill09,ying11}). These signatures of temperature-dependent flow velocities can be better explored by 1D models using non-uniform heating function along the loop. The potential of velocity diagnosis will be further utilized in the future.

\acknowledgments
The authors thank the referee for his/her constructive comments. Y.L. thanks James Klimchuk for letting her use the new EBTEL code, and Dana Longcope and Wenjuan Liu for their help with EBTEL modeling and scientific discussions. Y.L. and M.D.D are supported by NSFC under grants 10878002, 10933003, ￼11373023, and NKBRSF under grant 2011CB811402. Y.L. is also supported by NSFC under grant 11103008 and CSC under file No. 2011619032. J.Q. is supported by NSF grant ATM-0748428. \textit{SDO} is a mission of NASA's Living With a Star Program. \textit{Hinode} is a Japanese mission developed and launched by ISAS/JAXA, collaborating with NAOJ as a domestic partner, and NASA (USA) and STFC (UK) as international partners. Scientific operation of the \textit{Hinode} mission is conducted by the \textit{Hinode} science team organized at ISAS/JAXA. Support for the post-launch operation is provided by JAXA and NAOJ (Japan), STFC (UK), NASA, ESA, and NSC (Norway).

\bibliographystyle{apj}

\end{document}